\documentclass{article} 
\usepackage{iclr2021_conference,times}

\usepackage{amsmath,amsfonts,bm}

\def\eqref#1{equation~\ref{#1}}

\def\1{\bm{1}}

\DeclareMathAlphabet{\mathsfit}{\encodingdefault}{\sfdefault}{m}{sl}
\SetMathAlphabet{\mathsfit}{bold}{\encodingdefault}{\sfdefault}{bx}{n}

\usepackage[hyphens]{url}           
\usepackage{hyperref}       

\usepackage{nicefrac} 

\usepackage{subfig}
\usepackage{graphicx}
\graphicspath{{images/}}

\usepackage{amsmath}
\usepackage{amsthm}

\newtheorem{theorem}{Theorem}

\newtheorem{assumption}{Assumption}

\title{Network Group Testing}

\author{Paolo Bertolotti and Ali Jadbabaie\\
Institute for Data, Systems, and Society\\
Massachusetts Institute of Technology\\
Cambridge, MA 02139, USA \\
\texttt{\{pbertolo,jadbabai\}@mit.edu}
}

\iclrfinalcopy 
\begin{document}

\maketitle

\begin{abstract}
We consider the problem of identifying infected individuals in a population of size $N$. We introduce a group testing approach that uses significantly fewer than $N$ tests when infection prevalence is low. The most common approach to group testing, Dorfman testing, groups individuals randomly. However, as communicable diseases spread from individual to individual through underlying social networks, our approach utilizes network information to improve performance. Network grouping, which groups individuals by community, weakly dominates Dorfman testing in terms of the expected number of tests used. Network grouping’s outperformance is determined by the strength of community structure in the network. When networks have strong community structure, network grouping achieves the lower bound for two-stage testing procedures. As an empirical example, we consider the scenario of a university testing its population for COVID-19. Using social network data from a Danish university, we demonstrate network grouping requires significantly fewer tests than Dorfman. In contrast to many proposed group testing approaches, network grouping is simple for practitioners to implement. In practice, individuals can be grouped by family unit, social group, or work group.
\end{abstract}

\section{Introduction}

Group testing improves testing capabilities for infectious diseases when resources are limited. In normal scenarios, infected individuals from a population of size $N$ are identified by testing all population members individually, which uses $N$ tests. In the simplest form of group testing, individual samples are pooled together into groups of size $n$ for an initial stage of testing. If a group tests negative, all individuals within the group are classified as negative for the disease. If a group tests positive, all individual samples from the group are retested individually to identify the infected members. To illustrate the power of group testing, consider the scenario where $N=50$ and one individual is infected. If individuals are pooled into groups of size $n=10$ for an initial stage of testing, one group will test positive and all 10 samples from the group will be retested. The group testing approach uses 15 tests compared to the 50 used under individual testing.

Group testing was introduced by \cite{dorfman1943} to screen for syphilis in the US military. Dorfman's insight was simple but powerful. As a result, group testing has been employed numerous times in the medical field for diseases including influenza, chlamydia, and malaria  \citep{van2012, currie2004, taylor2010}. Within the US, group testing is used in blood banks  and infertility prevention programs where large numbers of individuals are routinely tested \citep{fda_blood, bilder2010}. Group testing's efficient use of resources has made it a valuable technique in developing areas. Notably, group testing was used during the early stages of the HIV pandemic in Africa when polymerase chain reaction (PCR) test costs were high \citep{emmanuel1988}. By reducing testing costs and increasing access to diagnostic information, group testing plays an important role in increasing health equity.

Under Dorfman's approach, each individual's infection probability is treated as homogenous and individuals are placed into groups randomly, which is equivalent to ignoring any information regarding an individual's susceptibility to infection. However, as communicable diseases spread from individual to individual through underlying social networks, an individual's network location affects their infection probability. In this work, we utilize network information to pool individuals for group testing. Specifically, we group individuals by community as infections are more likely to spread between closely connected community members than between members of distinct communities.

In order to analyze the performance of a network grouping strategy, we introduce a generative network model and epidemic model. We derive the number of tests used under network grouping and prove the expected number of tests is upper bounded by Dorfman testing, which implies network grouping weakly dominates Dorfman. The outperformance of network grouping is determined by the strength of community structure in the network. In networks with strong community structure, network grouping performs optimally and achieves the lower bound for two-stage testing procedures. In networks with no structure, network grouping is equivalent to Dorfman testing.

We end by considering the scenario of a university testing its population for COVID-19 cases. Using social network data from a Danish university, we demonstrate network grouping outperforms Dorfman testing. Our work reinforces the benefit of group testing for communicable diseases, which is consequential for the current COVID-19 pandemic. Multiple labs have demonstrated the efficacy of group testing for detecting the SARS-CoV-2 virus \citep{hogan2020, yelin2020} and several countries have implemented group testing to increase their testing capabilities \citep{fda_pool, wsj_china}. As testing resources still remain constrained \citep{wsj_jan}, we hope more institutions and governments will take advantage of the power of group testing.

Since Dorfman's work in 1943, numerous group testing approaches with strong performance have been introduced \citep{litvak1994, cheraghchi2012, ghosh2020}. However, the complexity of the proposed methods have limited their adoption in the medical field. As a result, Dorfman testing remains the most common approach to group testing in practice \citep{mcmahan2012, fda_eua}. Importantly, our proposed approach is simple for practitioners to implement. In practice, individuals can be grouped by family unit, social group, work group, or other community structure.

\section{Setup}

In this section, we describe Dorfman two-stage testing and the lower bound for two-stage group testing procedures. Under two-stage testing, a population of size $N$ is split into $\nicefrac{N}{n}$ groups of size $n$ for an initial stage of testing. Let $G$ denote the number of positive groups after the initial stage. In the second stage of testing, all $n$ samples from each positive group are retested individually. In total, $\nicefrac{N}{n} + nG$ tests are used.

Under Dorfman testing, one individual is infected with probability one and the remaining $N-1$ individuals are infected independently with probability $v$.\footnote{In Dorfman's original paper, all $N$ individuals are infected independently with probability $v$. We deviate slightly from his original setup to ensure at least one individual is infected.} The expected number of infected individuals is therefore $\mathrm{E}[I_D] = 1 + (N-1)v$. The expected number of tests used under Dorfman testing is
\begin{align}
\mathrm{E}[T_D] = \frac{N}{n} + n\left[1 + \left(\frac{N}{n} - 1\right)v'\right]
\label{ET_D}
\end{align}
where $v' = 1-(1-v)^n$. The derivation of $\mathrm{E}[T_D]$ was provided by Dorfman and can be found in appendix A.1 for completeness. When the infection prevalence $v$ is low, Dorfman testing uses significantly fewer than $N$ tests in expectation. As an example, consider the scenario where ${N=1000}$ and $v=0.05$ (5\%). If we employ Dorfman testing and a group size of $n=10$, only 507 tests are needed in expectation to test the entire population, a reduction of nearly 50\% compared to the $N=1000$ tests used under individual testing. 

Given a population, a certain number of infected individuals, and a group size, the minimum number of tests is achieved by minimizing the number of positive groups $G$. $G$ is minimized by perfect grouping, in which all infected individuals are pooled together into the minimum possible number of groups. The lower bound for two-stage testing procedures when $1+(N-1)v$ individuals are infected is
\begin{align}
T_{LB} = \frac{N}{n} + n \cdot \max \left( 1,  \frac{1+(N-1)v}{n} \right)
\label{T_LB}
\end{align}
The derivation can be found in appendix A.2. Revisiting our example, if $N=1000$, $v=0.05$, and $n=10$, the minimum number of tests needed under two-stage group testing is 151. The lower bound is unattainable in most scenarios as we do not know which samples are infected a priori.

\section{Model}

In this work, we consider the population of $N$ individuals to be embedded in a network, where each individual corresponds to a node and their physical interactions correspond to edges. In our framework, the network underlying the population is generated by a stochastic block model (SBM). Specifically, we consider an SBM with $N$ nodes split into $\nicefrac{N}{m}$ communities of size $m$. Within each community of $m$ nodes, edges exist between nodes independently with probability $p$. Edges exist between nodes in different communities independently with probability $q$, where $q \leq p$. As a result, nodes are more likely to be connected to other nodes in the same community than to nodes in other communities.

For our epidemic model, we consider the initial stage of a branching process model. Specifically, an epidemic starts with a single infected seed node, which is chosen at random from the population. The seed node infects each of its neighbors independently with probability $\alpha$. The seed node has $m-1$ possible neighbors within its community, each connected with probability $p$, and $N-m$ possible neighbors outside of its community, each connected with probability $q$. As a result, the expected number of infected individuals under this model, which will we use for network grouping, is $\mathrm{E}[I_{NG}] = 1+(m-1)p\alpha + (N-m)q\alpha$. 

The epidemic model describes the initial stage of an outbreak or, alternatively, a super-spreader event. We set $\alpha$ such that the expected number of infected individuals in the epidemic model is equal to the expected number of infected individuals in the Dorfman setting. Setting $\mathrm{E}[I_{NG}] = \mathrm{E}[I_{D}]$ and solving for $\alpha$ yields $\alpha = (N-1)v / [(m-1)p + (N-m)q]$. For the remainder of this work, we assume the following.
\begin{assumption}
Assume $1 \leq n \leq N$, $1 < m < N$, $0 \leq q \leq p \leq 1$, and $v \in [0,1]$. In addition, assume $\alpha \in [0,1]$ set such that $\mathrm{E}[I_{NG}] = \mathrm{E}[I_{D}]$.
\label{assumption1}
\end{assumption}

\section{Results}
In this section, we introduce our main results regarding network grouping and its performance compared to Dorfman testing. Under network grouping, we group individuals by community. In the simplest case, if communities have the same size as groups, $m=n$, each community is pooled into a unique group. If community size is divisible by group size, the $m$ community members are pooled into $\nicefrac{m}{n}$ groups. If group size is divisible by community size, each group of size $n$ consists of $\nicefrac{n}{m}$ communities. For example, if $m=20$ and $n=10$, each community is pooled into two groups and if $m=5$ and $n=10$, each group consists of two communities. When $m$ not divisible by $n$ and $n$ not divisible by $m$, we keep communities intact as much as possible and remainder community members are pooled into the remaining groups.

The expected number of tests used under network grouping is 
\begin{align}
\mathrm{E}[T_{NG}] = \frac{N}{n} + n\left[1 + \left(\frac{m}{n} - 1\right)^+p' + \left(\frac{N}{n} - 1 - \left(\frac{m}{n} - 1\right)^+\right)q' \right]
\label{ET_NG}
\end{align}
where $p' = 1-(1-p\alpha)^n$, $q' = 1-(1-q\alpha)^n$, and $(x)^+ = \max(x,0)$. The derivation of the full distribution of the number of tests is provided in appendix A.3.

With the number of tests under network grouping, Dorfman testing, and the lower bound derived, we come to the main result of our work. The expected number of tests under network grouping is upper bounded by Dorfman testing and lower bounded by the two-stage testing lower bound.

\begin{theorem}
\label{theorem1}
Under the conditions of assumption \ref{assumption1}, $\mathrm{E}[T_{NG}]$ is increasing in $q$ and
\begin{align}
T_{LB} \leq \mathrm{E}[T_{NG}] \leq \mathrm{E}[T_D]
\end{align}
If $q=0$ and $n \geq m$, then $\mathrm{E}[T_{NG}] = T_{LB}$. \\
If $q=p$, then $\mathrm{E}[T_{NG}] = \mathrm{E}[T_D]$.
\end{theorem}
The proof of theorem 1 can be found in appendix A.4. Theorem 1 states network grouping weakly dominates Dorfman testing in terms of the expected number of tests used. The outperformance of network grouping is driven by $q$, the probability an edge exists between nodes in different communities. In settings where networks have extremely strong community structure, network grouping performs optimally and achieves the lower bound. Specifically, when $q=0$, communities are disconnected from each other and all infected individuals will reside within the same community. When $n \geq m$, each group is large enough to capture each entire community and, as a result, all infected individuals will be grouped together. However, there are also scenarios where network grouping is equivalent to Dorfman testing, notability when $q=p$. Interestingly, even though we assume a network model, epidemic model, and network grouping, we end up back where we started with Dorfman testing. The reasoning is simple: since the network has no structure, all nodes have the same probability of being infected and the network provides no useful information for grouping.

\begin{figure}[!t]
\centering
\subfloat[]{\includegraphics[width=2.75in, keepaspectratio]{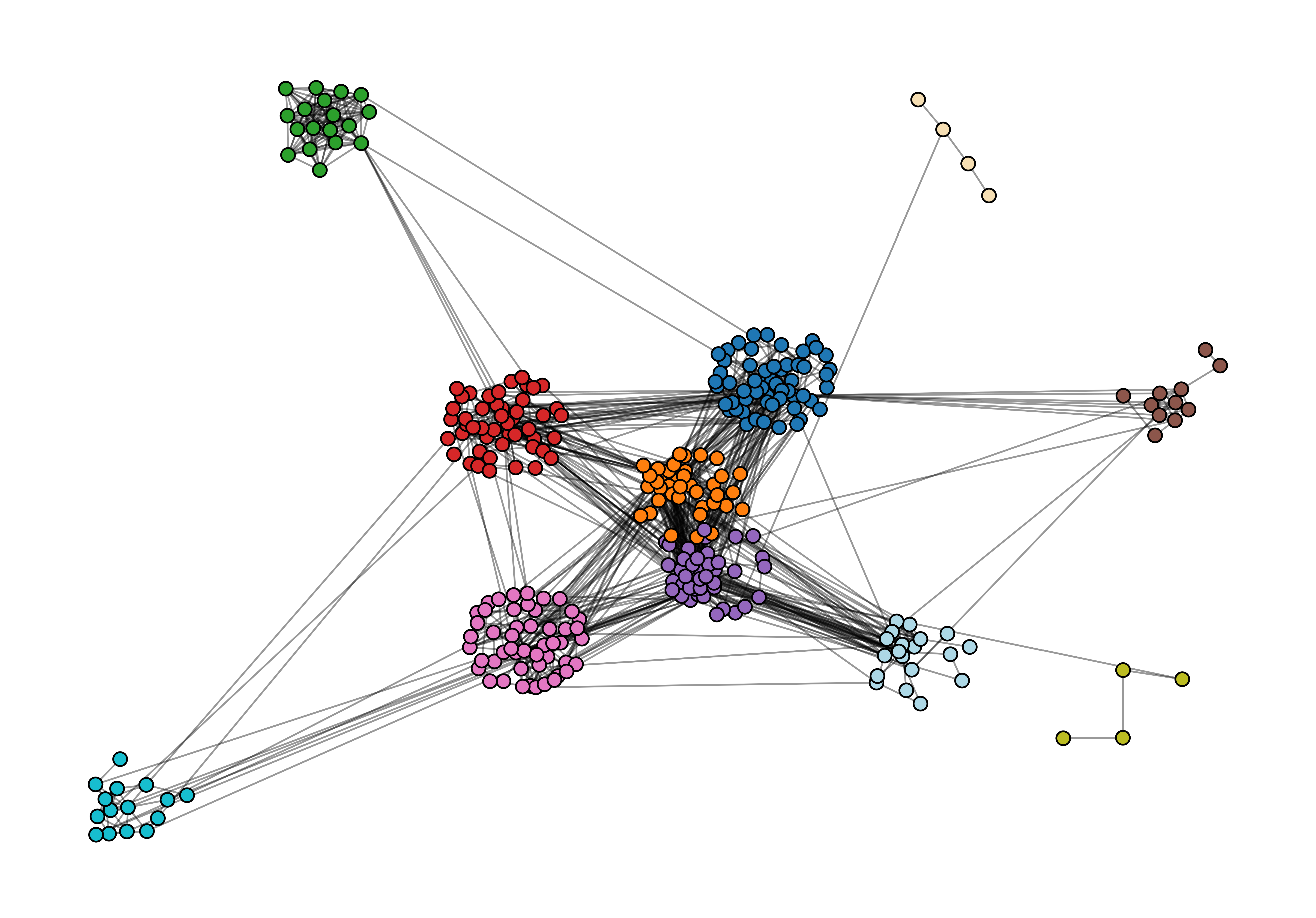}}%
\subfloat[]{\includegraphics[width=2.75in, keepaspectratio]{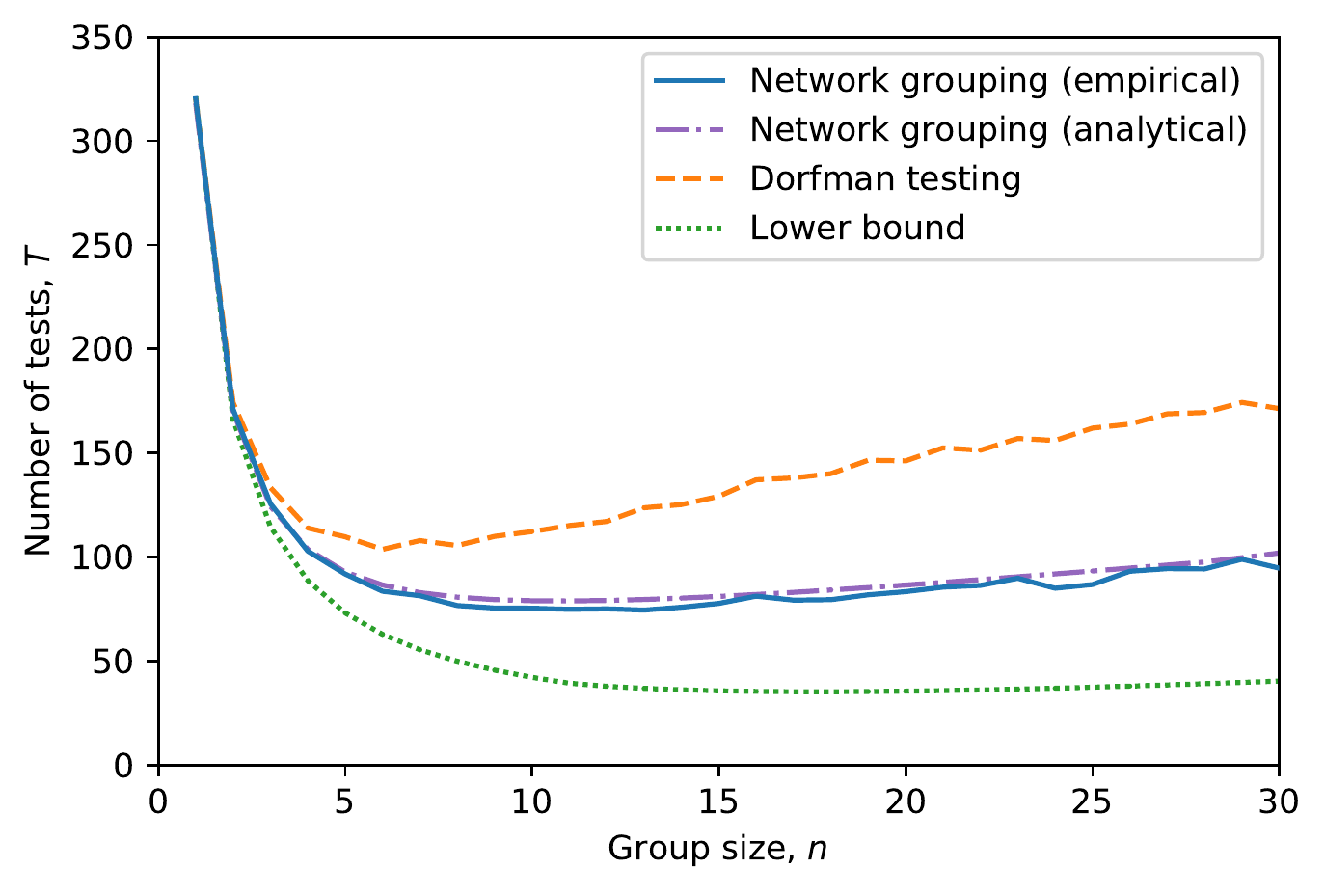}}%
\caption{\textbf{(a)} Social network of student interactions from the Technical University of Denmark. Nodes correspond to first-year students at the university and edges correspond to their physical interactions. Nodes are colored by their community affiliation, which is determined using the Louvain algorithm. \textbf{(b)} Comparison of testing approaches applied to the Danish university social network. The figure displays the average number of tests used to screen the population of $N=310$ as a function of group size. Averages are computed after simulating 1000 epidemic processes per group size.} 
\label{copenhagen_network}
\end{figure}

In normal cases when $0 < q < p$, network grouping significantly outperforms Dorfman testing. Consider the scenario of a university testing its population for COVID-19 cases. Using data from the Copenhagen Networks Study of \cite{sapiezynski2019}, we build the social network of first-year students at the Technical University of Denmark. The network contains 310 nodes, which correspond to students, and 1503 edges, which correspond to their physical interactions recorded using bluetooth-enabled smartphones. We apply the Louvain algorithm of \cite{blondel2008} to detect communities, resulting in 11 communities with an average size of 28 individuals. The social network with nodes colored by their community is displayed in figure \ref{copenhagen_network}a. We estimate $p$ to be 0.18 and $q$ to be 0.01, indicating a sparse network with strong community structure.

We simulate 1000 epidemic processes on the network using the model outlined in section 3 with $\alpha=0.95$, resulting in an estimated infection prevalence of 0.03 (3\%). We apply network grouping, which groups individuals by community, and Dorfman testing, which groups individuals randomly. Figure \ref{copenhagen_network}b displays the average number of tests used under the different approaches as a function of group size. Network grouping strongly outperforms Dorfman testing. When $n=10$, Dorfman testing uses 112 tests on average while network grouping uses 75 tests to screen the population of 310 students, a reduction of 33\%. In addition, figure \ref{copenhagen_network}b demonstrates our analytical result for the number of tests used under network grouping, provided in equation \ref{ET_NG}, is a strong approximation for the number of tests used in a real network setting.

In this work, we have introduced the idea of using social network information to improve group testing. When networks have strong community structure, network grouping outperforms Dorfman testing in terms of the number of tests used. It turns out network grouping also outperforms Dorfman in terms of false positives and false negatives when tests are imperfect. However, we leave discussion of imperfect tests to future write-ups. Importantly, network grouping is simple for practitioners to implement; individuals can be grouped by family unit, friend group, or other community structure.

\section*{Acknowledgments}
We thank Alberto Abadie, Jordan Brooks, Ben Deaner, Yash Deshpande, David Hughes, Peter Kempthorne, and Noelle Wyman for helpful comments and conversations. We are grateful to Eric Lai and the NIH RADx team for helpful discussions. The authors acknowledge the MIT SuperCloud and Lincoln Laboratory Supercomputing Center for providing high performance computing resources. Paolo Bertolotti was supported by a National Defense Science and Engineering Graduate (NDSEG) Fellowship.

\bibliography{bibliography}
\bibliographystyle{iclr2021_conference}

\newpage
\appendix
\section{Appendix}

\subsection*{A.1 Derivation of Dorfman testing}

Under Dorfman testing, a population of size $N$ is split into $\nicefrac{N}{n}$ groups of size $n$ for an initial stage of testing. Let $G$ denote the number of positive groups after the initial stage. In the second stage of testing, all $n$ samples from each positive group are retested individually. In total, $\nicefrac{N}{n} + nG$ tests are used. $G$ is a random variable. Of the $\nicefrac{N}{n}$ groups, one is positive with probability one as there is at least one infected individual. The remaining $\nicefrac{N}{n} - 1$ groups are positive independently with some probability $v'$. As a result, $G$ is distributed $1 + Bin(\nicefrac{N}{n} - 1, v')$.

The probability $v'$ is derived as follows. Each of the remaining $N-n$ individuals (that are not in the first group) are infected with probability $v$ and not infected with probability $1-v$. The probability that all $n$ individuals in a group are not infected is $(1-v)^n$. The probability that at least one individual in the group is infected, and therefore the group tests positive, is $v' = 1 - (1-v)^n$. Putting everything together, the number of tests used under Dorfman testing is distributed
\begin{align}
T_D \sim \frac{N}{n} + n\left[1 + Bin\left( \frac{N}{n} - 1, v' \right)\right]
\label{T_D}
\end{align}
Taking the expectation of $T_D$ provides $\mathrm{E}[T_D]$ as displayed in equation \ref{ET_D}.

\subsection*{A.2 Derivation of the two-stage lower bound}

Under two-stage group testing, a population of size $N$ is split into $\nicefrac{N}{n}$ groups of size $n$ for an initial stage of testing. Let $G$ denote the number of positive groups after the initial stage. In the second stage of testing, all $n$ samples from each positive group are retested individually. In total, $\nicefrac{N}{n} + nG$ tests are used. Given a population, a certain number of infected individuals, and a group size, the minimum number of tests is achieved by minimizing the number of positive groups $G$. $G$ is minimized by perfect grouping, in which all infected individuals are pooled together into the minimum possible number of groups.

When $1+(N-1)v$ individuals are infected, the minimum number of positive groups of size $n$ is $[1+(N-1)v] / n$. For example, if 20 individuals are infected and $n=10$, the minimum number of positive groups is two. When the number of infected individuals is greater than or equal to one but less than or equal to $n$, the minimum number of positive groups will be one. Note, there is always at least one infected individual in our framework. Putting everything together, the lower bound for the number of tests needed under two-stage group testing is 
\begin{align}
T_{LB} = \frac{N}{n} + n \cdot \max \left( 1,  \frac{1+(N-1)v}{n} \right)
\end{align}

\subsection*{A.3 Derivation of network grouping}

Under two-stage group testing, a population of size $N$ is split into $\nicefrac{N}{n}$ groups of size $n$ for an initial stage of testing. Let $G$ denote the number of positive groups after the initial stage. In the second stage of testing, all $n$ samples from each positive group are retested individually. In total, $\nicefrac{N}{n} + nG$ tests are used. $G$ is a random variable.

The network contains $\nicefrac{N}{m}$ communities of size $m$. We consider cases where $n$ divisible by $m$ or $m$ divisible by $n$. First consider the case where $m \leq n$. Since we group individuals by community (as described at the beginning of section 4), the infected seed node and its $m-1$ community members will be contained in the same group. This group will be positive with probability one. The remaining $\nicefrac{N}{n} - 1$ groups each contain $n$ nodes that belong to different communities than the seed node. As a result, each node in the remaining $\nicefrac{N}{n} - 1$ groups is not infected with probability $1-q\alpha$, as they are only infected if they are both connected to the seed, with probability $q$, and infected by the seed, with probability $\alpha$. The probability all $n$ nodes within a group are not infected is $(1-q\alpha)^n$. The probability that at least one individual in a group is infected, and therefore the group tests positive, is $q' = 1 - (1-q\alpha)^n$. In summary, the remaining $\nicefrac{N}{n} - 1$ groups are positive independently with probability $q'$. Putting everything together, the distribution of the number of tests used under network grouping when $m \leq n$ is
\begin{align}
T_{NG} \sim \frac{N}{n} + n\left[1 + Bin\left( \frac{N}{n} - 1, q' \right)\right]
\end{align}

Now consider the case where $m > n$. As we group individuals by community, there will be one group that contains the infected seed node and $n-1$ of its community members. This group will be positive with probability one. The remaining $m-n$ nodes from the seed node's community will be pooled into $(m-n)/n = \nicefrac{m}{n} - 1$ other groups. Each node in these groups will be infected with probability $1-p\alpha$, as they are only infected if they are both connected to the seed, with probability $p$, and infected by the seed, with probability $\alpha$. Following the same logic as the $m \leq n$ case, each of these $\nicefrac{m}{n} - 1$ groups is positive independently with probability $p' = 1 - (1-p\alpha)^n$. After accounting for the infected seed's group and the other $\nicefrac{m}{n} - 1$ groups, $\nicefrac{N}{n} - \nicefrac{m}{n}$ groups still remain. Each of the $n$ nodes in these groups are members of different communities than the seed node. Therefore, each of the $\nicefrac{N}{n} - \nicefrac{m}{n}$ groups is positive independently with probability $q' = 1 - (1-q\alpha)^n$. Putting everything together, the distribution of the number of tests used under network grouping when $m > n$ is
\begin{align}
T_{NG} \sim \frac{N}{n} + n\left[1 + Bin\left(\frac{m}{n} - 1, p' \right) + Bin\left( \frac{N}{n} - \frac{m}{n}, q' \right)\right]
\end{align}

The two cases, $m \leq n$ and $m > n$, can be easily combined. Defining $(x)^+ = \max(x,0)$, we have $(\nicefrac{m}{n}-~1)^+~=~0$ when $m \leq n$. Therefore, we can write the distribution of the number of tests used under network grouping in the general case as 
\begin{align}
T_{NG} \sim \frac{N}{n} + n\left[1 + Bin\left(\left(\frac{m}{n} - 1\right)^+, p' \right) + Bin\left( \frac{N}{n} - 1 - \left(\frac{m}{n} -1\right)^+, q' \right)\right]
\label{T_NG}
\end{align}
where $p' = 1-(1-p\alpha)^n$ and $q' = 1-(1-q\alpha)^n$. Equation \ref{T_NG} holds exactly when $N$ divisible by $n$ and either $m$ divisible by $n$ or $n$ divisible by $m$. It is a strong approximation otherwise. Taking the expectation of $T_{NG}$ in equation \ref{T_NG} provides $\mathrm{E}[T_{NG}]$ as displayed in equation \ref{ET_NG}.

\subsection*{A.4 Proof of theorem 1}

\paragraph*{Upper bound} To prove $\mathrm{E}[T_{NG}] \leq \mathrm{E}[T_D]$, we prove $\mathrm{E}[T_{NG}]$ is increasing in $q$ under assumption \ref{assumption1} and equals $\mathrm{E}[T_D]$ when $q$ is set to its maximum value under assumption \ref{assumption1}, $q=p$. To prove $\mathrm{E}[T_{NG}]$ is increasing in $q$, we consider the cases where $m > n$ and $m \leq n$ separately.

\textit{Case 1:} We first consider the case where $m > n$. When $m > n$, $\mathrm{E}[T_{NG}]$ is given by
\begin{align}
\mathrm{E}[T_{NG}] = \frac{N}{n} + n\left[1 + \left(\frac{m}{n} - 1\right)p' + \left(\frac{N}{n}  - \frac{m}{n} \right)q' \right]
\label{case1}
\end{align}
where $p' = 1-(1-p\alpha)^n$ and $q' = 1-(1-q\alpha)^n$. Under assumption \ref{assumption1}, ${\alpha = (N-1)v / [(m-1)p + (N-m)q]}$. Taking the derivative of equation \ref{case1} with respect to $q$ and simplifying yields
\begin{align}
\frac{\partial \, \mathrm{E}[T_{NG}]}{\partial q} = \frac{n v (N-1) (N-m) \left[p(m-1)\left(1 - q\alpha \right)^{n-1}-p (m-n)\left(1- p \alpha \right)^{n-1}\right]}{[(m-1)p + (N-m)q]^2}
\label{partial1}
\end{align}
Demonstrating equation \ref{partial1} is nonnegative proves equation \ref{case1} is weakly increasing in $q$. The denominator is nonnegative due to the square and $n$, $v$, $N-1$, and $N-m$ are nonnegative by assumption \ref{assumption1}. Examining the bracket term in the numerator, we note $p(m-1) \geq p (m-n)$ as $n \geq 1$ and $(1 - q\alpha)^{n-1} \geq (1- p \alpha)^{n-1}$ as $p \geq q$. Note both $1 - q\alpha$ and $1 - p\alpha$ are probabilities between 0 and 1 as $p$ and $\alpha$ are between 0 and 1. As a result, the bracket term is nonnegative and the entirety of equation \ref{partial1} is nonnegative.

\textit{Case 2:} We now consider the case where $m \leq n$. When $m \leq n$, $\mathrm{E}[T_{NG}]$ is given by
\begin{align}
\mathrm{E}[T_{NG}] = \frac{N}{n} + n\left[1 + \left(\frac{N}{n}  - 1 \right)q' \right]
\label{case2}
\end{align}
where $q' = 1-(1-q\alpha)^n$. Again, $\alpha = (N-1)v / [(m-1)p + (N-m)q]$. Taking the derivative of equation \ref{case2} with respect to $q$ and simplifying yields
\begin{align}
\frac{\partial \, \mathrm{E}[T_{NG}]}{\partial q} = \frac{ n v p  (N-1) (N-n) (m-1) \left(1 - q \alpha\right)^n}
{[(m-1)p +(N-m)q] \, [(m-1)p + (N - m)q + q(1-N)v]}
\label{partial2}
\end{align}
All terms in the numerator and the first bracket term in the denominator are nonnegative by assumption \ref{assumption1}. The second bracket term in the denominator is nonnegative if
\begin{align}
(m-1)p + (N - m)q + q(1-N)v &\geq 0
\label{cond1}
\end{align}
Rearranging equation \ref{cond1} yields
\begin{align}
\frac{q(N-1)v}{(m-1)p  +(N-m)q} &\leq 1 \\
q\alpha &\leq 1
\end{align}
which is true by assumption \ref{assumption1}. As a result, equation \ref{partial2} is nonnegative and equation \ref{case2} is weakly increasing in $q$.

\textit{Final step:} We have shown $\mathrm{E}[T_{NG}]$ is increasing in $q$ for $m > n$ and $m \leq n$. Setting $q$ to its maximum value under assumption \ref{assumption1}, $q=p$, we have $p' = q'$ as $1-(1-p\alpha)^n = 1-(1-q\alpha)^n$. In addition, $\alpha$ simplifies to $\nicefrac{v}{p}$ and $p\alpha = v$. Therefore, $p' = q' = v'$ where $v' = 1-(1-v)^n$. $\mathrm{E}[T_{NG}]$ in the general case simplifies to 
\begin{align}
\mathrm{E}[T_{NG}] &= \frac{N}{n} + n\left[1 + \left(\frac{m}{n} - 1\right)^+p' + \left(\frac{N}{n} - 1 - \left(\frac{m}{n} - 1\right)^+\right)q' \right] \\
&= \frac{N}{n} + n\left[1 + \left(\frac{N}{n} - 1\right)v' \right] 
\end{align}
and we have $\mathrm{E}[T_{NG}] = \mathrm{E}[T_{D}]$, completing the upper bound portion of the proof.

\paragraph*{Lower bound} To prove $\mathrm{E}[T_{NG}] \geq T_{LB}$, we prove $\mathrm{E}[T_{NG}] - T_{LB} \geq 0$ for the three cases of 1) group size larger than (or equal to) the expected number of infected individuals, $n \geq 1+(N-1)v$,  2) group size less than infected individuals and less than community size, $n < 1+(N-1)v$ and $n < m$, and 3) group size less than infected individuals and greater than (or equal to) community size, $n < 1+(N-1)v$ and $n \geq m$.

\textit{Case 1:} When group size is larger than or equal to the expected number of infected individuals, $n \geq 1+(N-1)v$, the lower bound in equation \ref{T_LB} simplifies to $\nicefrac{N}{n} + n$. Therefore,
\begin{align}
\mathrm{E}[T_{NG}] - T_{LB} = n\left(\frac{m}{n} - 1\right)^+p' + n\left(\frac{N}{n} - 1 - \left(\frac{m}{n} - 1\right)^+\right)q'
\label{lowercase1}
\end{align}
where $p' = 1-(1-p\alpha)^n$ and $q' = 1-(1-q\alpha)^n$. By assumption \ref{assumption1}, $n \geq 1$ and both $p'$ and $q'$ are probabilities between 0 and 1, as $1 - p\alpha$ and $1 - q\alpha$ are between 0 and 1. In addition, the term $(\nicefrac{N}{n} - 1 - (\nicefrac{m}{n} - 1)^+)$ is nonnegative as $N \geq n$ and $N > m$ . As a result, the entirety of equation \ref{lowercase1} is nonnegative.

\textit{Case 2:} When group size is smaller than the expected number of infected individuals and community size, $n < 1+(N-1)v$ and $n < m$, we can write the lower bound $T_{LB}$ as
\begin{align}
T_{LB} &= \frac{N}{n} + 1+(N-1)v \\
&= \frac{N}{n} + 1+(m-1)p\alpha + (N-m)q\alpha      \label{lowercase2_tlbinitial} \\
&=\frac{N}{n} + n +(m-1)(1-(1-p\alpha)) + (N-m)(1-(1-q\alpha)) - (n - 1)  \label{lowercase2_tlb}
\end{align}
where the second equality makes use of the $E[I_D] = E[I_{NG}]$ equivalence specified in assumption \ref{assumption1}. Equation  \ref{lowercase2_tlb} is a slight rearrangement of equation \ref{lowercase2_tlbinitial}. As $n < m$, $\mathrm{E}[T_{NG}]$ becomes
\begin{align}
\mathrm{E}[T_{NG}] &= \frac{N}{n} + n + (m - n)p' + (N-m)q' \\
&= \frac{N}{n} + n + (m - 1)p' + (N-m)q' - (n-1)p'  \label{lowercase2_etng}
\end{align}
Subtracting equation \ref{lowercase2_tlb} from equation \ref{lowercase2_etng} yields 
\begin{align}
\mathrm{E}[T_{NG}] &- T_{LB} = \nonumber \\
 (m &- 1)[p' -(1-(1-p\alpha))] +(N - m)[q' - (1-(1-q\alpha))]+(n - 1)(1-p')
\label{lowercase2}
\end{align}
By assumption \ref{assumption1}, we have $m > 1$, $N>m$, and $n \geq 1$. In addition, $1 \geq p'$ as $p' = 1-(1-p\alpha)^n$ is a probability between 0 and 1. Lastly, $p' = 1-(1-p\alpha)^n \geq 1-(1-p\alpha)$ as $(1-p\alpha)^n \leq (1-p\alpha)$. Similarly, $q' = 1-(1-q\alpha)^n \geq 1-(1-q\alpha)$. As a result, $\mathrm{E}[T_{NG}] - T_{LB}$ is nonnegative.

\textit{Case 3:} We consider the case where group size is smaller than the expected number of infected individuals but larger than (or equal to) community size, $n < 1+(N-1)v$ and $n \geq m$. Using equation \ref{lowercase2_tlbinitial} and the inequality $m \geq 1+(m-1)p\alpha$, we have the following inequality for the lower bound $T_{LB}$.
\begin{align}
T_{LB} &= \frac{N}{n} + 1+(m-1)p\alpha + (N-m)q\alpha   \\
&\leq \frac{N}{n} + m + (N-m)q\alpha \\
&= \frac{N}{n} + n + (N-m)(1-(1-q\alpha)) - (n-m)
\label{lowercase3_tlb}
\end{align}
Since $n \geq m$, $\mathrm{E}[T_{NG}]$ becomes
\begin{align}
\mathrm{E}[T_{NG}] &= \frac{N}{n} + n + \left(N - n \right)q'     \\
&= \frac{N}{n} + n + \left(N - m \right)q' - (n-m)q'
\label{lowercase3_etng}
\end{align}
Subtracting equation \ref{lowercase3_tlb} from equation \ref{lowercase3_etng} yields
\begin{align}
\mathrm{E}[T_{NG}]  - T_{LB} \geq (N-m)[q'-(1-(1-q\alpha))] + (n-m)(1-q')
\end{align}
By assumption, $N >m$ and $n \geq m$. In addition, $1 \geq q'$ as $q'$ is a probability between 0 and 1. Lastly, $q' = 1-(1-q\alpha)^n \geq 1-(1-q\alpha)$ as $(1-q\alpha)^n \leq (1-q\alpha)$. As a result, the difference $\mathrm{E}[T_{NG}]  - T_{LB}$ is nonnegative.

We have proven $\mathrm{E}[T_{NG}] - T_{LB} \geq 0$ for the three cases under consideration, completing the lower bound portion of the proof. \qed

\paragraph*{Final statements} The statement "If $q=0$ and $n \geq m$, then $\mathrm{E}[T_{NG}] = T_{LB}$" follows directly from the definition of $\mathrm{E}[T_{NG}]$. When $q=0$, we have ${q' = 1-(1-q\alpha)^n = 0}$. When $n \geq m$ and $q' = 0$, $\mathrm{E}[T_{NG}]$ simplifies to
\begin{align}
\mathrm{E}[T_{NG}] = \frac{N}{n} + n
\end{align}
By assumption \ref{assumption1}, $\alpha = (N-1)v / [(m-1)p + (N-m)q] \leq 1$. When $q=0$, we have ${(N-1)v / [(m-1)p] \leq 1}$, which implies $v \leq (m-1)p / (N-1)$. Recall the number of infected individuals is $1+(N-1)v$. We now have $1+(N-1)v \leq 1+(m-1)p$ and  $ 1+(m-1)p \leq m$ as $p \leq 1$. Since $m \leq n$, we have $1+(N-1)v \leq n$. Therefore, the lower bound $T_{LB}$ is
\begin{align}
T_{LB} = \frac{N}{n} + n
\end{align}
and $\mathrm{E}[T_{NG}] = T_{LB}$, completing the proof. \qed

Note, the assumption $\alpha \leq 1$ in assumption \ref{assumption1} sets an upper bound for the infection prevalence $v$ as $\alpha$ is a function of $v$. However, this is not restrictive as group testing is employed in cases when $v$ is low.

The statement "If $q=p$, then $\mathrm{E}[T_{NG}] = \mathrm{E}[T_D]$" is proved above during the upper bound portion of the proof.

\end{document}